Open Journal of Earthquake Research, 2021, 10, 17-29
https://www.scirp.org/journal/ojer
ISSN Online: 2169-9631
ISSN Print: 2169-9623

# The Theoretical and Practical Foundations of Strong Earthquake Predictability

Oleg Elshin[1], Andrew A. Tronin[2]

[1]President at Terra Seismic, Alicante, Spain/Baar, Switzerland
[2]Chief Scientist at Terra Seismic, Director at Saint-Petersburg Scientific-Research Centre for Ecological Safety of the Russian Academy of Sciences, St Petersburg, Russia
Email: oleg.elshin@terraseismic.com

## Abstract

Earthquakes and the tsunamis they produce are the world's most devastating natural disasters, affecting more than 100 countries. Not surprisingly, the problem of earthquake prediction has occupied scientists' minds for more than two thousand years. This paper provides theoretical and practical arguments regarding the possibility of predicting strong and major earthquakes worldwide. Many strong and major earthquakes can be predicted at least two to five months in advance, based on identifying stressed areas that begin to behave abnormally before strong events, with the size of these areas corresponding to Dobrovolsky's formula. We make predictions by combining knowledge from many different disciplines: physics, geophysics, seismology, geology, and earth science, among others. An integrated approach is used to identify anomalies and make predictions, including satellite remote sensing techniques and data from ground-based instruments. Terabytes of information are currently processed every day with many different multi-parametric prediction systems applied thereto. Alerts are issued if anomalies are confirmed by a few different systems. It has been found that geophysical patterns of earthquake preparation and stress accumulation are similar for all key seismic regions. The same earthquake prediction methodologies and systems have been successfully applied in global practice since 2013, with the technology successfully used to retrospectively test against more than 700 strong and major earthquakes since 1970. In other words, the earthquake prediction problem has largely been solved. Throughout 2017-2021, results were presented to more than 160 professors from 63 countries.

## 1. Introduction: Three Views on Earthquake Prediction

Humans have tried to predict earthquakes since the days of ancient Greece. Pherecydes of Syros made the first known forecast about 2500 years ago, predicting a strong earthquake after he discovered that the usually crystal-clear water from local wells had become dirty and undrinkable. Nowadays, thousands of scientists and dozens of scientific groups are working on this problem. Modern seismology and public opinion hold the view that earthquake prediction is currently impossible. Terra Seismic, which has been successfully making global earthquake predictions for many years, disagrees with this opinion [1]-[6]. With regard to this matter, we can divide the scientists involved into three groups:

The first—a large, rapidly growing number of people—believe that earthquake prediction is indeed possible, and are developing various earthquake forecasting methods. One of the best known is a family of M8 algorithms from Keilis-Borok's school that uses various kinds of seismic anomalies or deviations from "normal" seismic behavior [7] [8] [9]. Among other traditional approaches are electromagnetic methods [10] [11] [12], hydrological methods based on changes in the composition and levels of water [13] [14] [15], changes in the level of radon [16] [17], the observation of unusual clouds [18] [19], changes in meteorological data and animal behavior [20] etcetera. Over the past decade we have seen a dramatic growth in the new, primarily satellite remote sensing methods that study earthquakes [21] [22]. Among these methods are OLR, measurements of the earth's surface temperature, changes in the ionosphere [23], GPS [24] [25] [26] and other approaches [27] [28].

The second group think that whilst earthquake prediction might seem impossible today, it may become possible in the future once new knowledge has been accumulated or new phenomena discovered.

Finally, there is a third group of principal opponents who claim that earthquake prediction is inherently impossible [28] [29].

Based on our many years' experience in global earthquake prediction, in this paper we will offer a theoretical and practical explanation as to how real global earthquake prediction has been successfully undertaken by Terra Seismic on a global scale.

## 2. Discussion—Can Earthquakes Be Predicted?

The principal opponents put forward four key arguments (in italics below). They say:

*1) There are no reliable precursors, or those that there are prove to be too unreliable.*

We agree that precursors that are 100% reliable are rare. However, we have found effective solutions regarding how to deal with many unreliable precursors. Applying global Big Data methodologies and analyzing more than 700 strong and major earthquakes from the past, we identified dozens of new types of precursors or anomalies that appeared before strong events. Although each anomaly may occur in only 40 - 50 percent of total cases, if the area of future earthquakes is confirmed by 8 - 10 anomalies, the probability that they may simultaneously appear by chance for the same area is actually very small. Thus, by applying numerous new anomalies, we can make reliable predictions.

*2) The system is chaotic and therefore completely unpredictable. Stress can be randomly transferred across many fault lines and an earthquake can occur at any fault.*

We cannot agree with this either. For strong and major earthquakes, an enormous amount of stress has to be accumulated in the area of the future event. This massive stress cannot arise at one point from nowhere and a lengthy period of time is required for such an accumulation. The transfer of a huge amount of stress from one area to another is impossible since it cannot be explained through the fundamental laws of physics. In fact, we can "see" a linear stress accumulation in the area of a future earthquake. Our practical observations, therefore, contradict this theory of a chaotic system.

*3) The system is characterized by its self-organized criticality—a property of dynamical systems that have bifurcation points. The behavior in the vicinity of the point is characterized by the fact that with a small perturbation, the system can pass the bifurcation point, thereby completely changing its behavior model. The classic examples of self-organized criticality are the phase transition or sandpile model.*

There are a large number of different models and the applicability of each model to any particular system has to be reasonably justified. In particular, the sandpile model does not take into account the significant frictions that act between plates below the earth's surface etcetera. Furthermore, we cannot in fact be certain about the processes and state of matter located at depths of more than 15 - 20 km, where the stresses that cause most earthquakes are formed. Thus, we believe that this model is not applicable to earthquake preparation processes and that their conclusions are therefore incorrect.

*4) Nobody can predict earthquakes at present.*

We have been successfully predicting earthquakes worldwide since 2012 and our systems have been in practical use since 2013. Our technology has been used to retrospectively test data collected since 1970, with our systems successfully detecting about 90% of all significant earthquakes over the past 50 years. In other words, the earthquake prediction problem has largely been solved—see **Figures 2-11**. Throughout 2017-2021, results were presented to more than 160 professors from 63 countries.

## 3. Our Simple Model That Proves That Earthquakes Can Be Predicted

We propose a simplified model describing earthquake preparation—see **Figure 1**. We have two seismic systems with two plates. System A has two small adhesion zones between them that will subsequently lead to two small earthquakes.

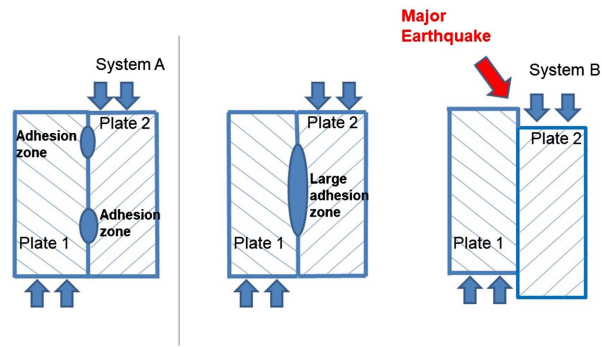

**Figure 1.** Model showing the generation of small and large earthquakes. A shows a system with a two small adhesion zones while B shows a system with a large adhesion zone.

In system B, there is one large adhesion zone that will cause a massive earthquake, with plates moving relatively to each other. We see two main forces in the process: forces driving the plates and the opposing forces of friction or adhesion between them. Obviously, the larger the adhesion zones between plates, the bigger the forces of friction between them. Based on classroom physics, we argue that due to the presence of a much larger adhesion zone in system B, this system will behave differently to system A. System B is "locked" and therefore will behave abnormally. The earth's surface above system B will be stressed, producing a numerous unusual phenomena or anomalies that can be detected before the event will strike.

Anyone who agrees that system B will behave differently than system A should thus accept that strong and major earthquakes can be predicted. Global earthquake prediction is about identifying areas around the world where B systems are currently forming. These areas of forthcoming strong and major earthquakes can be revealed by collecting, processing and analyzing huge amount of data, identifying anomalies and comparing current and historical data. We would expect that the amount of stress will gradually increase and the size of the observed stress area will increase as well as the event approaches. Each stressed area will also have its own "resistance limit" which, when reached, will see the strong event occur.

Based on this model and its implications, global earthquake prediction is possible and has been successfully implemented for many years. We have seen countless examples of the successful validation of this model in practice.

## 4. Practical Cases

We will illustrate our findings by following practical cases. See Figures 2-11.

## 5. Global Earthquake Prediction Systems—A Review of the Technology

Many strong and major earthquakes with magnitude M6.2 or greater can be predicted at least two to five months before they occur. Earthquake predictions are currently being provided for twenty-four key earthquake-prone regions:

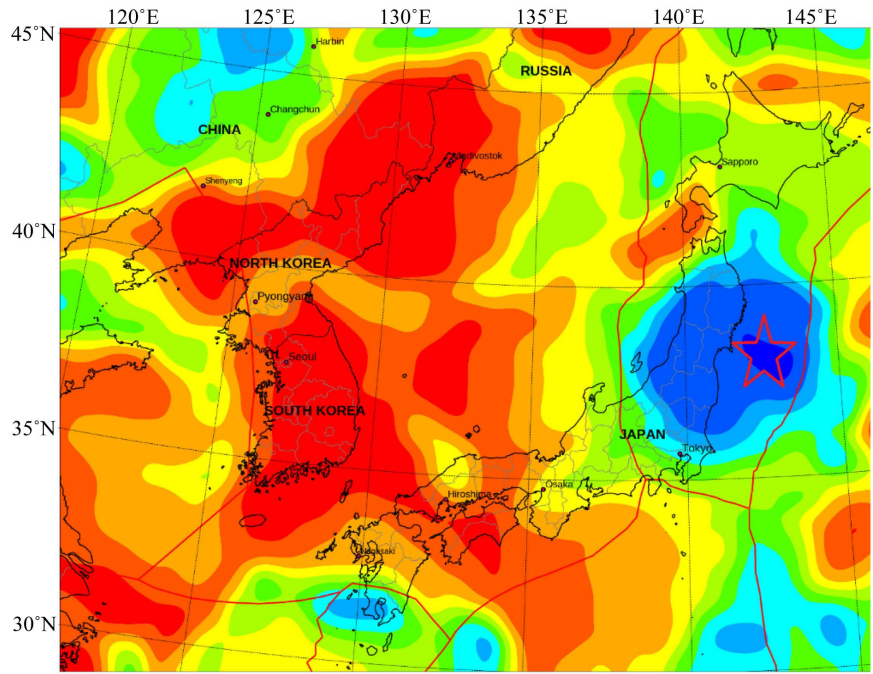

**Figure 2.** Historic testing of Terra Seismic systems for past famous seismic events. Stressed area in Japan before the 2011 Tohoku megaquake, shown in dark blue. Unstressed areas shown in other colors (red, green etc.).

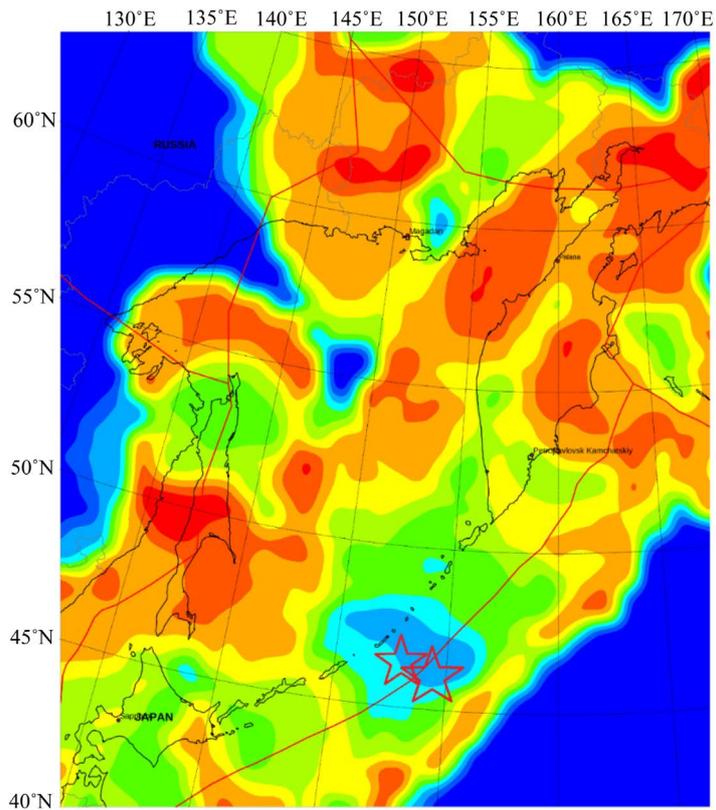

**Figure 3.** Historic testing of Terra Seismic systems for past famous seismic events. Stressed area in the Kuril Islands before the 2006-2007 two M8 earthquakes, shown in dark blue. Unstressed areas shown in other colors (red, green etc.).

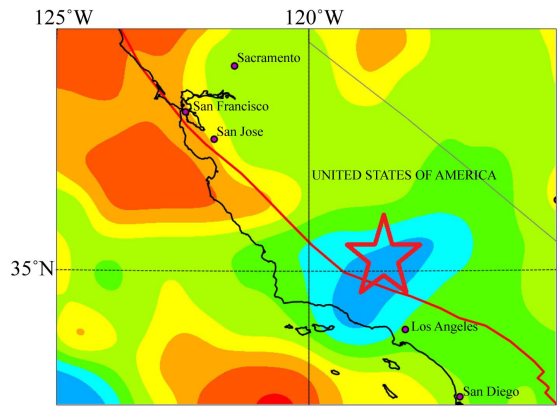

**Figure 4.** The case for a successfully predicted major earthquake. Stressed area in California before the 2019 Ridgecrest earthquake sequence, shown in dark blue.

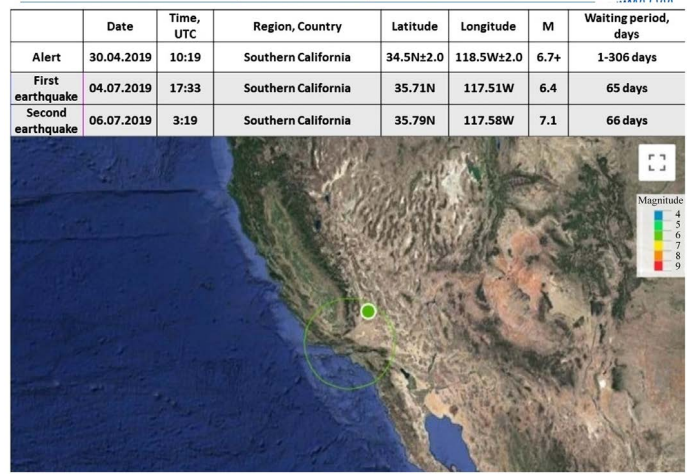

**Figure 5.** The case for a successfully predicted major earthquake. It shows a successful alert that was issued for the 2019 Ridgecrest earthquake sequence. The green circle indicates the prognostic signal.

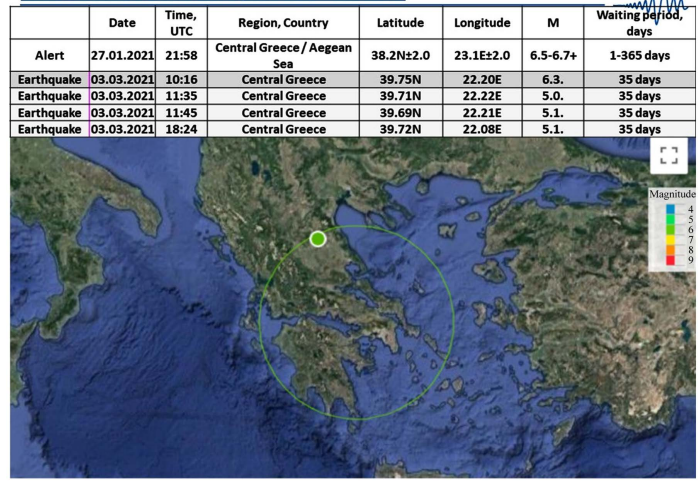

**Figure 6.** The case for a successfully predicted strong earthquake in Central Greece. The green circle indicates the prognostic signal and green dot shows the location of epicenter.

O. Elshin, A. A. Tronin

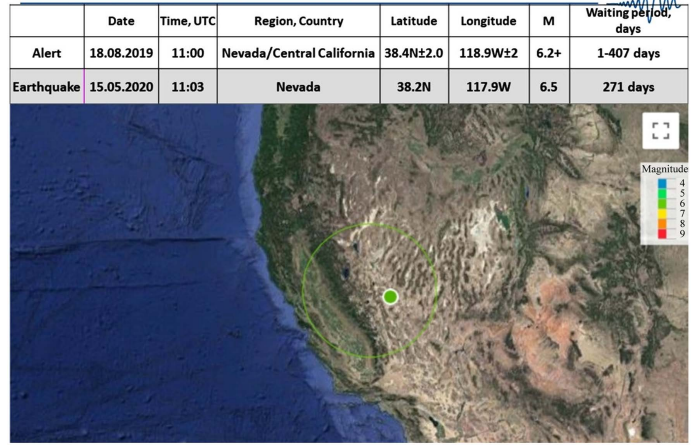

**Figure 7.** Successfully predicted strong earthquake in Nevada. The green circle indicates the prognostic area and green dot shows the location of epicenter.

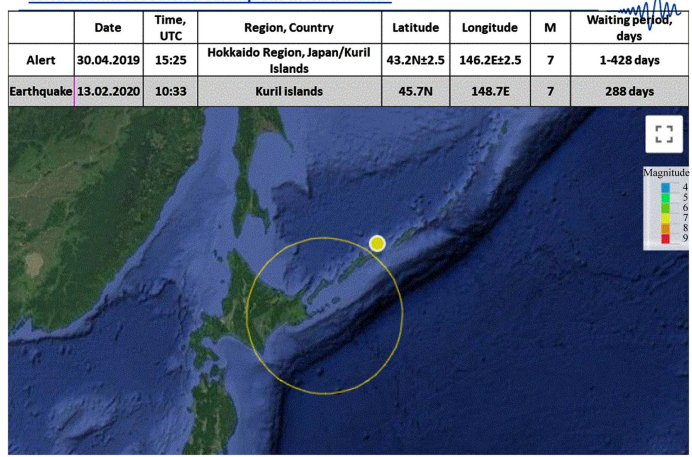

**Figure 8.** The case for a successfully predicted major earthquake in Kuril Islands. The yellow circle indicates the prognostic signal and yellow dot shows the location of epicenter.

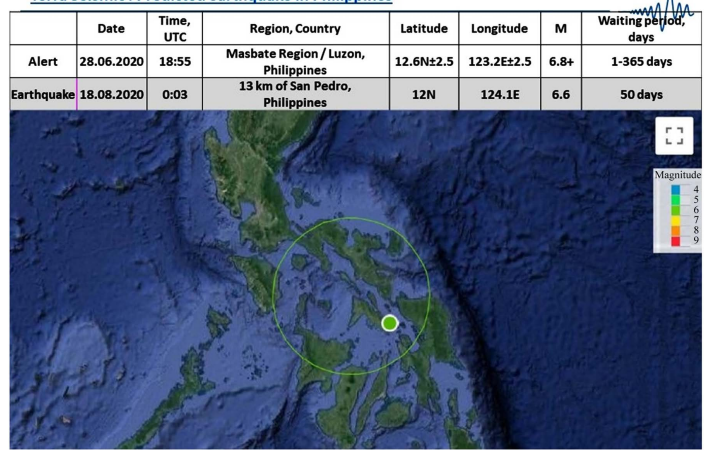

**Figure 9.** Successfully predicted strong earthquake in Philippines. The green circle indicates the prognostic area and green dot shows the location of epicenter.

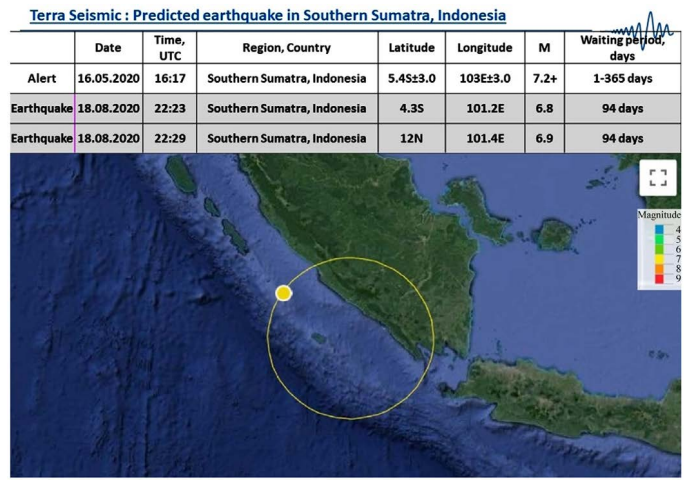

**Figure 10.** The case for a successfully predicted two strong earthquakes in Southern Sumatra, Indonesia. The yellow circle indicates the prognostic signal and yellow dot shows the location of epicenter.

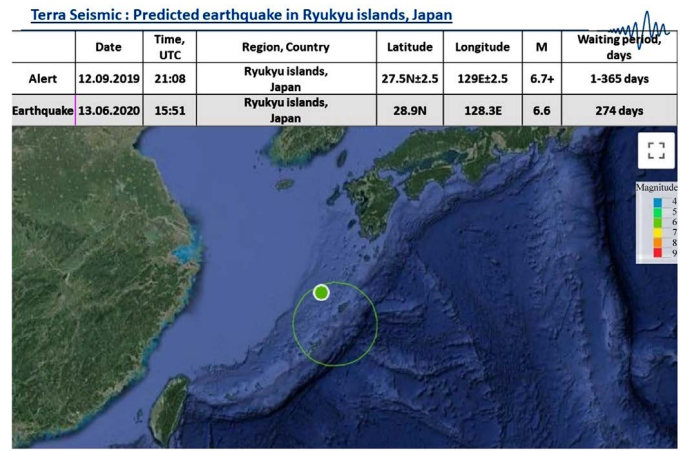

**Figure 11.** Successfully predicted strong earthquake in Ryukyu islands (Japan). The green circle indicates the prognostic area and green dot shows the location of epicenter.

Alaska, the Balkans, California, Canada, the Caribbean, Central America, Chile, China, Greece, India and Pakistan, Indonesia, Iran, Italy, the Izu Islands, Japan, Kamchatka and the Kuril Islands, Mexico, Central Asia, New Zealand, Okinawa, Papua New Guinea, Peru, Philippines, and Turkey.

The epicenter of a future strong or major earthquake can be determined with a high degree of confidence, currently as a circle with a radius of 150 - 250 km. Since the technology is permanently improving, it will be possible to locate epicenters more accurately. Generally, we do not promise to forecast a seismic event if the epicenter of the earthquake is located deeper than 30 km, although exceptions are possible. Fortunately, in many cases such deep earthquakes do not produce any significant damage since the energy of the earthquake dissipates before reaching the Earth's surface.

We develop long-term (from 2 to 5 years), mid-term (from 2 months to 2 years), and short-term (from 10 to 60 days) global prediction systems. The mid-term sys-tems are the most reliable and can predict most strong and major earthquakes at least 1 - 2 years in advance. The closer event to us—the most accurate assessment of magnitude and location we can make. A range of expected magnitudes for alerted events can be defined, for example, such as M6.5 - 6.8+. We are improving our systems on daily basis. After each new strong or major event occurs, we analyze it and include all new data to further fine tune our systems.

## 6. Methods: A Global Approach Based on the Collection of Worldwide Geophysical Data Online Combined with Historical Analysis

Our global earthquake prediction methodology is based on generally accepted assumptions—the discovering and revealing of the real geophysical processes that always occur prior to strong and major earthquakes. Earthquakes strike unexpectedly for humans, but they are not sudden for nature. In nature, earthquakes are the result of releasing accumulated stress, which is formed through the gradual geophysical processes that develop below earth's surface over long periods of time. To produce a strong or major earthquake, enormous stress needs to be accumulated. For example, a magnitude eight earthquake releases the same amount of energy as the joint explosion of about one thousand atomic bombs of the size dropped on Hiroshima. We argue that such enormous stress accumulation can be successfully detected well in advance. The area of the forthcoming earthquake will be stressed and will behave differently to unstressed areas.

In practice, a global earthquake prediction is very similar to a doctor's diagnosis of a patient with a medical problem, kidney disease for example, by collecting multiple analyses and without surgery. In our case, the whole world serves as the examined "patient". Just as disease is a deviation from normalcy in the human body because something is wrong, abnormalities in geophysical parameters indicate that something is wrong in this particular part of the world. Before a strong or major earthquake, we notice anomalies in the data we are collecting which signal that a strong event is approaching. Also, strong and major earthquakes are very rare events; they may occur in a specific area every forty or fifty years or more. Before such historically rare events we observe unique combinations of certain parameters in stressed regions. Stronger earthquakes take longer to prepare, and so they can be detected earlier. For example, the preparation for a M8 event in Japan can be detected at least three to five years before the event. Usually, for events with greater magnitudes, larger stressed preparation areas are observed, a phenomenon that approximately corresponds to the estimates from the Dobrovolsky formula [30]. Also, as expected, we found that when preparing for events of greater magnitude, the size of the stress area gradually increases over an observed build-up period.

On a daily basis we develop and use the innovative Big Data satellite technology capable of processing and analyzing terabytes of information for all key seismically prone regions in almost real time. By using a global approach and historically accumulated and analyzed data, we have managed to significantly increase the amount of reliable data for developing and testing completely new theories, models and systems. For our systems, we have applied all available data on earthquakes, their precursors and working forecasting methods from all sources available over the past 2000 years.

## 7. Conclusions

Global earthquake prediction technology consists of identifying a seismically hazardous block that behaves in an anomalous manner before an event. Such a block can be expected to generate a strong earthquake with a magnitude greater than 6. The selection of such anomalous blocks is carried out by various methods and processing data available for a particular region using remote sensing methods, GPS methods, electromagnetic, meteorological and seismological methods which can all be applied in the process. The essence of the technology is to integrate signals from different methodologies, eliminate noise and identify such anomalous blocks worldwide. We make predictions by combining knowledge from many different disciplines: physics, geophysics, seismology, geology, and earth science, among others.

Every year, our planet records numerous seismic events: in an average year, there are about 100 - 120 M6 quakes, 10 - 15 M7 quakes and 1 - 2 M8 quakes. However, about 85 percent of these strong or major events occur in remote or sparsely populated areas and will actually produce no significant damage. Each earthquake prediction requires months of hard work. At Terra Seismic we, therefore, do not seek to predict all strong and major earthquakes on the planet simply because it makes no sense. From a practical point of view, humanity needs just about 15 - 20 of potentially dangerous earthquakes to be predicted every year. Our criteria for such potentially impact earthquakes: they are M6.2 - 6.5+ earthquakes (for low seismic areas) and 6.5 - 7.0+ earthquakes (for highly seismic areas), close to populated areas and with an epicenter depth of less than 30 km. Last year, 12 potentially dangerous earthquakes were predicted in different regions around the world. Our top priority is to help governments to save human lives and thus we call on all governments and agencies responsible for disaster preparedness management to unite efforts. We are also open to collaborate with all scientists working in this field.

## Acknowledgements

This project was not possible without the scientific data provided by different government agencies, international organizations, science institutions and academia.

We wish to thank US Geological Survey (USGS), European-Mediterranean Seismological Centre (EMSC), Japanese Meteorological Agency (JMA), National Aeronautical and Space Administration (NASA), National Oceanic and Atmospheric Administration (NOAA), European Space Agency (ESA), International

GNSS Service (IGS), Jet Propulsion Laboratory (JPL)/Caltech, Ionospheric Prediction Service (IPS), Weather Underground and World Data Center (WDC) in Kyoto, Japan.

## Conflicts of Interest

The authors declare no conflicts of interest regarding the publication of this paper.

## References

[1] Elshin, O. and Tronin, A.A. (2020) Global Earthquake Prediction Systems. *Open Journal of Earthquake Research*, **9**, 170-180. https://doi.org/10.4236/ojer.2020.92010

[2] Oleg Elshin, Focus GEOmedia 3-2020—With Terra Seismic Earthquake Prediction, We Can Be Better Prepared for Earthquakes in Italy.
https://geomediaonline.it/en/focus/344-with-terra-seismic-earthquake-prediction-we-can-be-better-prepared-for-earthquakes-in-italy

[3] Kalvik, J. (2019) Nobel Prize to Oleg Elshin and Terra Seismic Will Help Protect Humanity from Earthquakes and Tsunamis.
http://www.etterretningen.no/2019/08/16/nobel-prize-to-oleg-elshin-and-terra-seismic-will-help-protect-humanity-from-earthquakes-and-tsunamis

[4] Marr, B. (2015) FORBES: Big Data: Saving 13,000 Lives a Year by Predicting Earthquakes?
http://www.forbes.com/sites/bernardmarr/2015/04/21/big-data-saving-13000-lives-a-year-by-predicting-earthquakes

[5] CNN Philippines (2019) Live Interview with Oleg Elshin.
https://www.youtube.com/watch?v=x1erNJVtM4U&fbclid=IwAR19kZFjwZgiA3B6bP_3hmF_uJ3m2_p3YC3uGJIfZ1Uyop7uybDYVkU9E

[6] Marr, B. (2016) Big Data in Practice. How 45 Successful Companies Used Big Data Analytics to Deliver Extraordinary Results. Wiley, Hoboken.
https://doi.org/10.1002/9781119278825

[7] Keilis-Borok, V.I. and Kossobokov, V.G. (1990) Premonitory Activation of Earthquake Flow: Algorithm M8. *Physics of the Earth and Planetary Interiors*, **61**, 73-83.
https://doi.org/10.1016/0031-9201(90)90096-G

[8] Shebalin, P., Kellis-Borok, V., Gabrielov, A., Zaliapin, I. and Turcotte, D. (2006) Short-Term Earthquake Prediction by Reverse Analysis of Lithosphere Dynamics. *Tectonophysics*, **413**, 63-75. https://doi.org/10.1016/j.tecto.2005.10.033

[9] International Commission on Earthquake Forecasting for Civil Protection (ICEF) (2011) Operational Earthquake Forecasting: State of Knowledge and Guidelines for Utilization. *Annals of Geophysics*, **54**, 315-391.

[10] Hayakawa, M. (2019) Seismo Electromagnetics and Earthquake Prediction: History and New Directions. *International Journal of Electronics and Applied Research*, **6**, 1-23. https://doi.org/10.33665/IJEAR.2019.v06i01.001

[11] Varotsos, P., Alexopoulos, K., Nomicos, K., *et al.* (1986) Earthquake Prediction and Electric Signals. *Nature*, **322**, 120. https://doi.org/10.1038/322120a0

[12] De Santis, A., Marchetti, D., Pavón-Carrasco, F.J., *et al.* (2019) Precursory Worldwide Signatures of Earthquake Occurrences on Swarm Satellite Data. *Scientific Reports*, **9**, Article No. 20287. https://doi.org/10.1038/s41598-019-56599-1

[13] Barberio, M.D., Barbieri, M., Billi, A., *et al.* (2017) Hydrogeochemical Changes be-

fore and during the 2016 Amatrice-Norcia Seismic Sequence (Central Italy). *Scientific Reports*, **7**, Article No. 11735. https://doi.org/10.1038/s41598-017-11990-8

[14] Rikitake, T. (2001) Prediction and Precursors of Major Earthquakes. Terra Scientific, Tokyo, 197.

[15] Kopylova, G. and Boldina, S. (2020) Hydrogeological Earthquake Precursors: A Case Study from the Kamchatka Peninsula. *Frontiers in Earth Science*, **8**, Article ID: 576017. https://doi.org/10.3389/feart.2020.576017

[16] Hauksson, E. (1981) Radon Content of Groundwater as an Earthquake Precursor: Evaluation of Worldwide Data and Physical Basis. *Journal of Geophysical Research*, **86**, 9397-9410. https://doi.org/10.1029/JB086iB10p09397

[17] Hall, S.S. (2011) At Fault? *Nature*, **477**, 264-269. https://doi.org/10.1038/477264a

[18] Guo, G.M. and Jie, Y.Y. (2013) Three Attempts of Earthquake Prediction with Satellite Cloud Images. *Natural Hazards and Earth System Sciences*, **13**, 91-95. https://doi.org/10.5194/nhess-13-91-2013

[19] Tan, X., *et al.* (2014) Modeling Pre-Earthquake Cloud Shape from Remote-Sensing Images. 2014 *Third International Workshop on Earth Observation and Remote Sensing Applications* (*EORSA*), Changsha, 11-14 June 2014, 470-474. https://doi.org/10.1109/EORSA.2014.6927935

[20] Ikeya, M. (2004) Earthquakes and Animals. World Scientific, Singapore, 285. https://doi.org/10.1142/5382

[21] Tronin, A.A. (2010) Satellite Remote Sensing in Seismology. A Review. *Remote Sensing*, **2**, 124-150. https://doi.org/10.3390/rs2010124

[22] Rathje, E., Eeri, M. and Adams, B. (2008) The Role of Remote Sensing in Earthquake Science and Engineering: Opportunities and Challenges. *Earthquake Spectra*, **24**, 471. https://doi.org/10.1193/1.2923922

[23] Elliott, J., Walters, R. and Wright, T. (2016) The Role of Space-Based Observation in Understanding and Responding to Active Tectonics and Earthquakes. *Nature Communications*, **7**, Article No. 13844. https://doi.org/10.1038/ncomms13844

[24] Liu, J.Y., Chen, Y.I., Chen, C.H. and Hattori, K. (2010) Temporal and Spatial Precursors in the Ionospheric Global Positioning System (GPS) Total Electron Content Observed before the 26 December 2004 M9.3 Sumatra-Andaman Earthquake. *Journal of Geophysical Research*, **115**, A09312. https://doi.org/10.1029/2010JA015313

[25] Freund, F. (2010) Toward a Unified Solid State Theory for Pre-Earthquake Signals. *Acta Geophysica*, **58**, 719-766. https://doi.org/10.2478/s11600-009-0066-x

[26] Murray, J.R. and Svarc, J. (2017) Global Positioning System Data Collection, Processing, and Analysis Conducted by the U.S. Geological Survey Earthquake Hazards Program. *Seismological Research Letters*, **88**, 916-925. https://doi.org/10.1785/0220160204

[27] Cenni, N., Viti, M. and Mantovani, E. (2015) Space Geodetic Data (GPS) and Earthquake Forecasting: Examples from the Italian Geodetic Network. *Bollettino di Geofisica Teorica ed Applicata*, **56**, 129-150.

[28] Geller, R.J., Jackson, D.D., Kagan, Y.Y. and Mulargia, F. (1997) Earthquakes Cannot Be Predicted. *Science*, **275**, 1616. https://doi.org/10.1126/science.275.5306.1616

[29] Sykes, L.R., Shaw, B.E. and Scholz, C.H. (1999) Rethinking Earthquake Prediction. *Pure and Applied Geophysics*, **155**, 207-232. https://doi.org/10.1007/s000240050263

[30] Dobrovolsky, I.P., Zubkov, S.I. and Miachkin, V.I. (1979) Estimation of the Size of

Earthquake Preparation Zones. *Pure and Applied Geophysics*, **117**, 1025-1044.
https://link.springer.com/article/10.1007/BF00876083
https://doi.org/10.1007/BF00876083